\documentclass[%
 amsmath,amssymb,
aps,
pra,
notitlepage]{revtex4-1}
\usepackage{graphicx}

\begin{document}
\title{A General Approach to Casimir Force Problems Based on Local Reflection Amplitudes and Huygen's Principle.}
\author{C. D. Markle and R. Cowsik}
\affiliation{McDonnell Center for the Space Sciences. Physics Department,\\ Washington University in St. Louis}
\date{\today}
\begin{abstract}
In this paper we describe an approach to Casimir Force problems that is ultimately generalizable to all fields, boundary conditions, and cavity geometries. This approach utilizes locally defined reflection amplitudes to express the energy per unit area of any Casimir interaction. To demonstrate this approach we solve a number of Casimir Force problems including the case of uniaxial boundary conditions in a parallel-plate cavity.
\end{abstract}
\maketitle
\section{Introduction}
Since Casimir's pioneering work in 1948 many advancements have taken place in the field of Casimir force theory and its experimental confirmation. Casimir's original work described the force that operates between uncharged parallel conductors arising from the vacuum energy of the electromagnetic field. The theory has since been extended to isotropic dielectrics\cite{Lifshitz,Dzyaloshinskii,Lifshitz2} and anisotropic materials\cite{Parsegian,Barash}. Work in related fields\cite{Ji,Ji2} may require further generalization leading to new and interesting results. Experimentation in the field began in the 1950's with the efforts of Overbeek and Sparnaay\cite{Sparnaay} and of Derjaguin and Abrikosova\cite{Derjaguin} and has continued to the present day with ever increasing precision\cite{Kim,Munday2,Chan,Decca}, covering a wide range of separations from $10\rm{nm}- 1\mu \rm{m}$.

However, 63 years after the birth of this field of study, many problems still remain. These problems include the modelling of material properties, the effects of temperature, and the approach to complex geometries among others. In addition to these out-standing problems, there exists another problem: the lack of a common approach. From mode counting methods to Green's functions to stress-energy tensor formalisms, each constitutes a unique method for calculating the Casimir force using exclusive nomenclature leading to unique interpretations of the phenomenon. Although most of these approaches yield the same or similiar predictions, it is often difficult to know exactly how to use any particular formalism to solve a new problem or assess the domain of its applicability. In an effort to redress this issue a number of excellent books have been written that detail many of these approaches \cite{Milloni,Milton,Bordag}. Herein we develop an approach that covers a wide-range of Casimir problems and we hope to describe it in detail so as to be easily adapted to new problems.

In this paper we will calculate the Casimir force using a scattering approach and subsequent mode counting method. Our basic formalism is not specific to any geometry or boundary conditions; we simply assume that we have two interacting boundaries as shown below and note that the right-bound waves in the cavity are generated by reflections of all the left-bound waves off the surface $S_1$ and vice-versa.
\begin{figure}[h]
\includegraphics[width=2.5in]{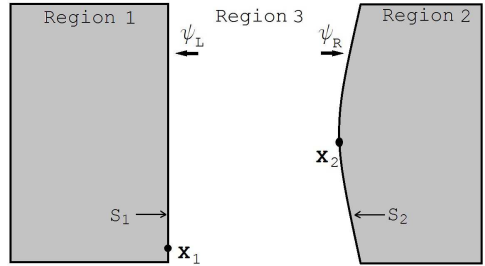}
\caption{\footnotesize{Two interacting surfaces $S_1$ and $S_2$ of arbitrary geometry. The field between the surfaces will be denoted as the sum of a right-bound wave and a left-bound wave.}\label{schematic}}
\end{figure}
The amplitude of the right-bound wave and left-bound wave at points $\mathbf{x}_1$ and $\mathbf{x}_2$ on surfaces $S_1$ and $S_2$ respectively are related by a generalized Huygen's Principle, namely, that the amplitude of any polarization state at a particular location is given by the sum of all incident wavelets added up in a phase sensitive manner:
\begin{equation}\label{R}
\Psi_R(\mathbf{x}_2)=\int_{S_1}g(\mathbf{x}_2,\mathbf{x}_1)\Gamma(\mathbf{x}_2,\mathbf{x}_1)R_1\Psi_{L}(\mathbf{x}_1) dS_1
\end{equation}
\begin{equation}\label{L}
\Psi_L(\mathbf{x}_1)=\int_{S_2}g(\mathbf{x}_1,\mathbf{x}_2)\Gamma(\mathbf{x}_1,\mathbf{x}_2)R_2\Psi_{R}(\mathbf{x}_2) dS_2
\end{equation}
Here $R_1$ and $R_2$ are $n\times n$ reflection matrices with dimension equal to the number of polarization states of the field under consideration (for the electro-magnetic field these are 2x2 matrices) and $g\Gamma$ equals the gradient of the Green's function. Specifically,
\begin{equation}\label{G}
\left(\frac{\partial G}{\partial n}\right)=g\Gamma
\end{equation}
where $\Gamma$, hereafter referred to as the transport functions or transport matrix, encodes only the phase changes and $g$ encodes only the modulus changes suffered by the wavelet during transport from some point $\mathbf{x}$ to some other point $\mathbf{x}'$. Both $g$ and $\Gamma$ are $n\times n$ matrices. The properties of Green's functions and Eq. \ref{G} requires $g$ to satisfy:
\begin{equation}
\int\int g(\mathbf{x},\mathbf{x'})g(\mathbf{x'},\mathbf{x''})dS  dS''=I
\end{equation}
Substituting equation 2 into equation 1 we get,
\begin{widetext}
\begin{eqnarray}
\Psi_{R}(\mathbf{x}_2)&=&\int_{S_1}\int_{S_2'}g(\mathbf{x}_2,\mathbf{x}_1)\Gamma(\mathbf{x}_2,\mathbf{x}_1)R_1g(\mathbf{x}_1,\mathbf{x'}_2)\Gamma(\mathbf{x}_1,\mathbf{x'}_2)R_2\Psi_{R}(\mathbf{x}_2') dS_1dS_2' \\
&=&\int_{S_1}\int_{S_2'}g(\mathbf{x}_2,\mathbf{x}_1)g(\mathbf{x}_1,\mathbf{x'}_2)\Gamma(\mathbf{x}_2,\mathbf{x}_1)R_1\Gamma(\mathbf{x}_1,\mathbf{x'}_2)R_2\Psi_{R}(\mathbf{x}_2') dS_1dS_2' \label{eigen1}
\end{eqnarray}
\end{widetext}
due to $g\Gamma Rg\Gamma R$ being invariant under cyclic permutation. The trivial solution to the above equation is $\Gamma R\Gamma R=0$ and will be disregarded. Thus the only non-trivial solutions to equation \ref{eigen1} is:
\begin{equation}\label{delta}
\Gamma(\mathbf{x}_2,\mathbf{x}_1)R_1\Gamma(\mathbf{x}_1,\mathbf{x}_2)R_2= I
\end{equation}

Equation \ref{delta} represents a system of equations whereby a non-trivial solution exists if and only if,
\begin{equation}\label{det}
\rm{det}(\rm{\textbf{I}}-\Gamma(\mathbf{x}_2,\mathbf{x}_1)R_1\Gamma(\mathbf{x}_1,\mathbf{x}_2)R_2)=0
\end{equation}
An eigenmode of the cavity exists for any choice of $\omega$, $\mathbf{k}$, $\mathbf{x}_1$, or $\mathbf{x}_2$ which can satisfy equation \ref{det}.

In order to calculate the Casimir interaction energy one must sum over the energies of all such eigenmodes, a task most easily performed by utilizing the argument principle:
\begin{equation}
\frac{1}{2}\hbar\sum_n\omega_n=\frac{\hbar}{4\pi i}\left[\int_{i\infty}^{-i\infty}\omega d\ln\Delta +\int_{C+}\omega d\ln\Delta\right]
\end{equation}
In the above expression $\Delta=\rm{det}(\rm{\textbf{I}}-\Gamma(\mathbf{x}_2,\mathbf{x}_1)R_1\Gamma(\mathbf{x}_1,\mathbf{x}_2)R_2)$. The transport functions, $\Gamma$, have elements proportional to $e^{-i\mathbf{k}\cdot\mathbf{r}}$, making the above integral oscillatory. To deal with this we can perform a change of variables, $\xi_c=-i\omega/c$. All the elements of $\Gamma$ will then be proportional to exponentially decaying functions and consequently the integral over the semicircle $C_+$ goes to zero. After integrating by parts the expression for the energy carried by all the modes in the cavity is given by:
\begin{equation}\label{E}
E=\frac{\hbar c}{4\pi}\sum_{\rm{paths}}\int_{-\infty}^{\infty} \rm{ln}(\rm{det}(\rm{\textbf{I}}-\Gamma(\mathbf{x}_2,\mathbf{x}_1)R_1\Gamma(\mathbf{x}_1,\mathbf{x}_2)R_2))d\xi_c
\end{equation}
where the sum is over all unobstructed straight line paths connecting the two surfaces.

\section{Comparison with previous results}
In this section we will demonstrate how to use this approach by solving a few specific cases and showing that this approach yields results consistent with past works. We will start with the simplest case, the Casimir case described by a parallel-plate cavity of infinite lateral extent $L_x=L_y=\infty$ and finite spacing $L_z=a$, the walls of which are perfect conductors, $R_1=R_2=\mathbf{I}$. The cavity itself is empty with $\epsilon=1$, a fact that allows us to say the transport functions are diagonal (no polarization mixing upon transport). Moreover, the transport of an eigenmode across the cavity is invariant under the change of polarization, this allows us to write the transport functions as: $\Gamma_{\mathbf{r}}=e^{i\mathbf{k}\cdot\mathbf{r}}\textbf{I}$, where $\mathbf{r}=(r_x,r_y,a)$ is the vector between $\mathbf{x}_1$ and $\mathbf{x}_2$. From this we can write equation \ref{E} as:
\begin{equation}\label{E2}
E=\frac{\hbar c}{4\pi}\int d\mathbf{k}\int d\mathbf{r}\int_{-\infty}^{\infty}\ln(\rm{det}(\textbf{I}-e^{-i\mathbf{k}\cdot\mathbf{r}}e^{-i\mathbf{k}'\cdot(-\mathbf{r})}\textbf{I}))d\xi_c
\end{equation}
Here $\mathbf{k}'=(k_\perp \cos\phi,k_\perp\sin\phi,-i\sqrt{\xi_c^2+k_\perp^2})$ and $\mathbf{k}=(k_\perp \cos\phi,k_\perp\sin\phi,i\sqrt{\xi_c^2+k_\perp^2})$ are the left and right bound wave vectors, respectively and the $d\mathbf{k}$ and $d\mathbf{r}$ integrals are a symbolic representation of the sum over all the connecting paths. We can express the integrand in a different form by noting that for any $2\rm{x}2$ matrix, $\mathbf{A}$, the following is true:
\begin{equation}\label{det2}
\rm{det}(\mathbf{I}-\mathbf{A})=1-\rm{tr}(\mathbf{A})+\rm{det}(\mathbf{A})
\end{equation}
Using equation \ref{det2} and the definitions of the wave-vectors we can rewrite equation \ref{E2} as:
\begin{equation}\label{ten}
E/A=2\frac{\hbar c}{4\pi}\int_0^\infty \frac{k_\perp dk_\perp}{2\pi}\int_0^{\infty}2\ln(1-e^{-2a\sqrt{\xi_c^2+k_\perp^2}})d\xi_c
\end{equation}
Here we have written $E/A$ as the energy per unit area in order to eliminate the integral over $d\mathbf{r}$ which symbolically represents an area integration the details of which will be addressed in the next section of this paper. We can directly evaluate the above equation by expressing the natural log as a series and performing the integration term by term:
\begin{eqnarray}
E/A&=&-\frac{\hbar c}{2\pi^2}\lim_{\delta^+ \rightarrow 0}\int_0^\infty k_\perp dk_\perp \int_{\delta}^\infty d\xi_c\sum_{n=1}^\infty \frac{1}{n}e^{-2an\sqrt{\xi_c^2+k_\perp^2}} \\
&=&-\frac{\hbar c}{2\pi^2}\lim_{\delta^+ \rightarrow 0}\sum_{n=1}^\infty \int_{0}^{\pi/2}\int_{\delta}^\infty\frac{e^{-2anp}}{n} p^2\sin\theta d\theta dp
\end{eqnarray}
here we have performed a change of variables: $\xi_c =p\cos\theta$ , $k_\perp =p\sin\theta$. The integration over $\theta$ can be carried out quickly to give a value of 1. The lower limit of integration of $p$ has been written as $\delta$ to eliminate any controversy over expanding the natural log around a discontinuity. Integrating over $p$ by parts and taking the limit as $\delta\rightarrow 0$ yields $1/(4a^3n^4)$, so that:
\begin{equation}
E/A=-\frac{\hbar c}{8\pi^2a^3}\sum_{n=1}^\infty\frac{1}{n^4}
\end{equation}
The sum in the above expression is well known and has a definite value of $\zeta(4)=\pi^4/90$. This yields the value for the energy per unit area:
\begin{equation}\label{casimir}
E_{cas}(a)=-\frac{\pi^2\hbar c}{720 a^3}
\end{equation}
in conformity with the standard Casimir result\cite{Milloni}.
\section{Parallel-plate cavity of finite size}
In the last section we calculated the Casimir energy per unit area of a parallel-plate cavity of infinite extent. Although this is an interesting result it is only applicable under the condition that the separation of the plates is much less than the size of the plates. Below we calculate the Casimir energy per unit area of a cavity of finite size.

\begin{figure}[h]
\includegraphics[width=1.5in]{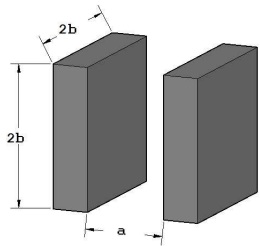}
\caption{\footnotesize{Two parallel plates separated by a distance $a$ in the z-direction. The plates are squares of side-length $2b$. The plates are aligned such that the bottom left corners of both plates are given by the xy-coordinates $(-b,-b)$ and the top right corners are given by $(b,b)$.}\label{finite}}
\end{figure}

For simplicity let us once again consider the case of a perfectly conducting parallel-plate cavity. This time however the side-walls will only occupy the region $(-b,b)$ in the x- and y-directions. The separation of the plates will once again be $a$ as shown in Figure \ref{finite}.
To solve this problem lets start with equation \ref{E2} and write the integral over $\mathbf{r}$ as an integral over either surface weighted by the probability that the set of isotropically distributed path vectors emanating from each differential area element is intercepted by the other surface.
\begin{eqnarray}\label{finiteeq}
E&=&\frac{\hbar c}{4\pi^2}\int\frac{(d\mathbf{A_1}\cdot\mathbf{\hat{r}})(d\mathbf{A_2}\cdot\mathbf{\hat{r}})}{\pi r^2}\int_0^{\infty}k_\perp dk_\perp\int_0^\infty\ln\Delta d\xi_c \nonumber \\
&=&\frac{-\pi ^2\hbar c}{720a^3}\frac{a^2}{\pi} \int_{\frac{-b}{a}}^{\frac{b}{a}}\frac{dx_1dy_1dx_2dy_2}{\left[\left(x_2-x_1\right)^2+\left(y_2-y_1\right)^2+1\right]^2} \label{plates}
\end{eqnarray}
The factor of $\pi$ in the denominator of the $\mathbf{r}$ integration is the normalization factor that is derived by noting that for a plate of unit area placed parallel to an infinite plane the spatial part of the integral should be unity, in order that Eq. \ref{finiteeq} agree with Eq. \ref{casimir} in the limit $b\rightarrow \infty$. Here we have performed the integration over the wave properties in the same manner as the preceding section and scaled the integration variables by $a$. From this we can see that Eq. \ref{plates} differs from Eq. \ref{casimir} solely in the manner in which the area is calculated.
\begin{figure}[h]
\includegraphics[width=3.5in]{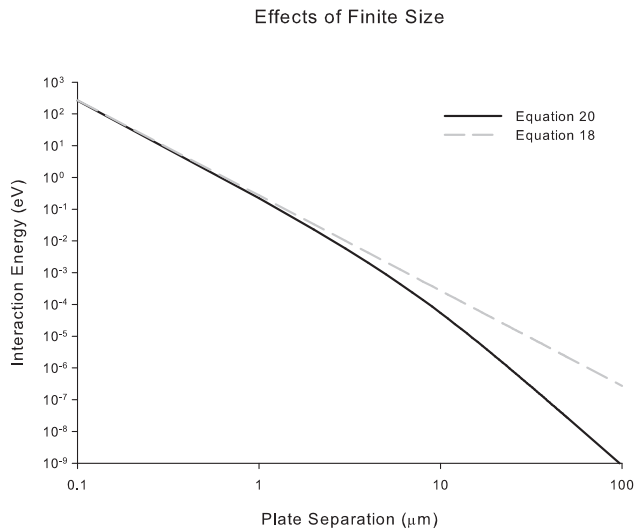}
\caption{\footnotesize{Log-Log graph of energy vs separation. This graph was generated for square plates of side-length 10 $\mu$m separated by distances between .1 $\mu$m and 100 $\mu$m. The dashed line represents Casimir's expression Eq. \ref{casimir} mulitplied by the area of Plate 1. The solid line represents our expression Eq. \ref{plates}.}\label{finite2}}
\end{figure}
To demonstrate the effects of finite size on the overall Casimir energy we have plotted the Casimir energy vs separation for square plates of side $2b=10\mu$m over separations of .1 $\mu$m to 100 $\mu$m. The dashed line in the graph was calculated using Casimir's expression Eq. \ref{casimir} simply multiplied by the area of Plate 1, the solid line was calculated using our expression Eq. \ref{plates}. From Figure \ref{finite2} it is apparent that for cavities with lateral dimension much larger than their separation ($b/a >> 1$) our expression approaches Casimir's expression, however for separations larger than the side-length ($b/a < 1$) our expression begins to drop off quicker than Casimir's expression. This behaviour is to be expected, as the plates should begin to appear as points rather than planes for large separation, leading to an additional $1/a^2$ behaviour.

\section{Parallel-plate Cavity described by uniaxial boundary conditions}
\begin{figure}[h]
\includegraphics[width=4in]{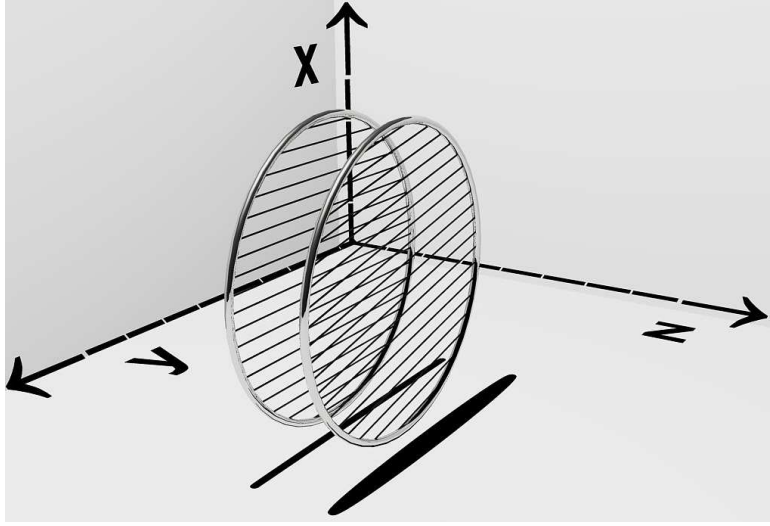}
\caption{\footnotesize{Two parallel plates separated by a distance $a$ in the z-direction. The coordinate axes are oriented to coincide with the principle axes of Plate 1 (left plate) with the x-direction coinciding with its optic axis. Plate 2 (right plate) is rotated about the z-axis by an angle $\chi$ relative to the Plate 1.}\label{schematic2}}
\end{figure}

An interesting case is that of a parallel-plate cavity with uniaxial boundary conditions as illustrated in Figure \ref{schematic2}. Such a cavity would have side-walls made of a uniaxial material, for example a wire grid polarizer or a graphite crystal cut with in-plane optical anisotropy. When two such plates are placed close together with their optic axes rotated by an angle $\chi$ with respect to each other, both an orientationally dependent normal force and a torque tending to align the optic axes of the plates have been predicted \cite{Barash}\cite{Munday}\cite{Shmuel}.
In order to address this Casimir problem in the formalism presented in this paper, let us start with equation \ref{E} relevant to a parallel-plate cavity of infinite extent:
\begin{widetext}
\begin{equation}\label{E/A}
E/A=\frac{\hbar c}{2\pi}\int d\mathbf{k}\int_0^\infty\ln(1-\rm{tr}(R_1R_2)e^{-2a\sqrt{\xi_c^2+k_\perp^2}}+\rm{det}(R_1R_2)e^{-4a\sqrt{\xi_c^2+k_\perp^2}})d\xi_c
\end{equation}
\end{widetext}
We need only to define $R_1$ and $R_2$ and proceed as before. In order to multiply them is this fashion these reflection matrices need to be in the same basis. We will write our reflection matrices in the TE-TM basis. Fortunately for all parallel-plate cavities the basis vectors that define the TE-TM basis for a given wave-vector on one surface are the same as for the other surface. Thus $R_1$ and $R_2$ will automatically be in the same basis. Were this not the case we would write $R_1=\mathcal{B}_{G\leftarrow L}\mathcal{R}_1\mathcal{B}_{L\leftarrow G}$, where $\mathcal{R}$ refers to the reflection matrix in a local coordinate basis. $L$ and $G$ refer to a local and global basis and $\mathcal{B}$ indicates a change of basis matrix which could be a function of position.

The reflection matrices are 2x2 matrices the elements of which describe transition amplitudes between the polarization states of the incident and reflected waves. 
\begin{eqnarray}
R_1=\left(\begin{array}{cc}
r_{1EE}&r_{1ME}\\
r_{1EM}&r_{1MM}
\end{array}\right)&,\quad &R_2=\left(\begin{array}{cc}
r_{2EE}&r_{2ME}\\
r_{2EM}&r_{2MM}
\end{array}\right)
\end{eqnarray}
Where $R_1$ and $R_2$ represent the reflection matrices of Plates 1 and 2 respectively. The functional definition of these matrices is to describe the reflected wave in terms of the incident wave. So, if we let $\psi_i=\left(1\ \ 0\right)$ represent an incident transverse electric wave with unit amplitude, we can write the reflected wave as $\psi_r=R_1\psi_i=\left(r_{1EE}\ \  r_{1EM}\right)$. 

In order to obtain an explicit expression for the matrices we must find the general solution for a wave in each of the three regions (the interior of the cavity and inside plates 1 and 2). We then have to match up these solutions with Maxwell's equations providing the boundary conditions.
In the region between the plates (Region 3) the general solution of Maxwell's equations is a linear combination of transverse plane waves described by an angular frequency $\omega$, wave vector $\mathbf{k}$, and polarization unit vector $\hat{\lambda}$. For any choice of $\omega$ we can write the solution as the sum of forward and backward travelling waves, each of which have two independent polarizations. Thus the solution for any of $\omega$ will be the sum of four waves, written in terms of the vector potential as:
\begin{equation}\label{soln3}
\mathbf{A}_3 = (a_3\hat{\lambda}_{E}+a_4\hat{\lambda}_{+M}) e^{i(\mathbf{k}\cdot \mathbf{r}-\omega t)} +(a_5\hat{\lambda}_{E}+a_6\hat{\lambda}_{-M}) e^{i(\mathbf{k}'\cdot \mathbf{r}-\omega t)}
\end{equation}
In the above equation $a_i$ is the amplitude of the $i^{\rm{th}}$ wave. The polarization and wave vectors can be written explicitly as:
\begin{equation}
\begin{array}{ccccccc}
\mathbf{k}=\left(\begin{array}{c} k_\perp \cos(\phi) \\k_\perp \sin(\phi) \\i\rho_3\end{array} \right) &,& \mathbf{k}' = \left(\begin{array}{c}k_\perp \cos(\phi) \\k_\perp\sin(\phi) \\-i\rho_3\end{array} \right) &,& \hat{\lambda}_{E}=\left(\begin{array}{c} \sin(\phi) \\-\cos(\phi) \\0 \end{array}\right) &,& \hat{\lambda}_{\pm M}=\frac{1}{\sqrt{\rho_3^2-k_\perp^2}}\left(\begin{array}{c} \cos(\phi)\rho_3 \\  \sin(\phi)\rho_3 \\ \pm ik_\perp\end{array} \right)
\end{array}
\end{equation}
In the above equations $k_\perp$ is the magnitude of the component of $\mathbf{k}$ perpendicular to the z-axis and $\rho_3=\sqrt{\epsilon_3\xi_c^2+k_\perp^2}$ where we have performed a scaled Wick rotation as before, $\xi_c=-i\omega/c$.

The general solutions in the interior of Plates 1 and 2 (Regions 1 and 2, respectively)\cite{Born} are similar but given in terms of a linear combination of ordinary and extraordinary waves denoted by the subscripts $o$ and $e$. In contrast with Region 3, the solutions in Region 1 and 2 will have two waves each instead of four because we assume Regions 1 and 2 to extend to infinity thereby eliminating any inbound waves. The solutions in Region 1 and 2 are then:
\begin{equation}
\begin{array}{ccc}\label{soln12}
\mathbf{A}_1 = a_1\hat{\lambda}_{1o} e^{i(\mathbf{k}_{1o}\cdot\mathbf{r}-\omega t)} + a_2\hat{\lambda}_{1e} e^{i(\mathbf{k}_{1e}\cdot\mathbf{r}-\omega t)} &,& \mathbf{A}_2 = a_7\hat{\lambda}_{2o} e^{i(\mathbf{k}_{2o}\cdot\mathbf{r}-\omega t)} + a_8\hat{\lambda}_{2e} e^{i(\mathbf{k}_{2e}\cdot\mathbf{r}-\omega t)}
\end{array}
\end{equation}
It's well known that the tangential components of the wave vector are left unchanged by refraction and as such, only the normal component of the wave vectors and the three components of the polarization vectors in the above expression need to be worked out explicitly. This can be done by solving the set of four coupled algebraic equations; (1): the norm of the polarization vector equals one ($\lambda ^2 = 1$), (2): the transversality of the wave ($\mathbf{k} \cdot \mathbf{D} = 0$), (3): the definitions of the ordinary ($\mathbf{D} \cdot \mathbf{\hat{O}} = 0$) and extra-ordinary waves($\mathbf{D}\cdot(\mathbf{k}_e\times\mathbf{\hat{O}}) = 0$), (4): Snell's law $\left(k_z = \pm i\sqrt{\xi_c^2\left(D^2/\mathbf{E}\cdot\mathbf{D} \right)+k_\perp^2}\right)$. Where $\mathbf{D}=\mathbf{\epsilon} \mathbf{E}$ is the electric displacement. The dielectric tensor for each plate is represented by:
\begin{equation}
\epsilon_1=\left(\begin{array}{ccc}\epsilon_{1\parallel}&0&0\\0&\epsilon_{1\perp}&0\\0&0&\epsilon_{1\perp}\end{array}\right)\quad ,\quad \epsilon_2=R_z ^T\left[\chi\right]\left(\begin{array}{ccc}\epsilon_{2\parallel}&0&0\\0&\epsilon_{2\perp}&0\\0&0&\epsilon_{2\perp}\end{array}\right)R_z\left[\chi\right]
\end{equation}
The resulting wave and polarization vectors can be written explicitly:

\begin{eqnarray}
k_{1oz}= -i\rho_1 = -i\sqrt{\epsilon_{1\perp}\xi_c^2+k_\perp^2} \qquad\qquad & k_{1ez}=-i\tilde{\rho}_1=-i\sqrt{\epsilon_{1\parallel}\xi_c^2+k_\perp^2+k_\perp^2\left(\frac{\epsilon_{1\parallel}}{\epsilon_{1\perp}}-1\right)\cos^2[\phi]}\\
& \nonumber \\
k_{2oz}= i\rho_2=i\sqrt{\epsilon_{2\perp}\xi_c^2+k_\perp^2} \qquad\qquad & k_{2ez}=i\tilde{\rho}_2=i\sqrt{\epsilon_{2\parallel}\xi_c^2+k_\perp^2+k_\perp^2\left(\frac{\epsilon_{2\parallel}}{\epsilon_{2\perp}}-1\right)\cos^2[\phi+\chi]}
\end{eqnarray}
\begin{eqnarray}
\hat{\lambda}_{1o}= \left(\begin{array}{c} 0\\ -\rho_1\\ik_\perp\sin[\phi]\end{array} \right) \qquad\qquad & \hat{\lambda}_{1e}= \left(\begin{array}{c} \epsilon_{1\perp}(k_\perp^2\sin^2[\phi]-\tilde{\rho}_1^2) \\k_\perp^2\epsilon_{1\parallel}\cos[\phi]\sin[\phi] \\ik_\perp\epsilon_{1\parallel}\cos[\phi]\tilde{\rho}_1  \end{array}\right) \qquad\qquad & \hat{\lambda}_{2o}= \left(\begin{array}{c} \sin[\chi]\rho_2\\ \cos[\chi]\rho_2  \\ ik_\perp\sin[\phi+\chi] \end{array}\right)
\end{eqnarray}
\begin{eqnarray}
\hat{\lambda}_{2e}= -\left(\begin{array}{c} k_\perp^2\sin[\phi+\chi](\epsilon_{2\parallel}\cos[\phi+\chi]\sin[\chi]-\epsilon_{2\perp}\cos[\chi]\sin[\phi+\chi])+\epsilon_{2\perp}\cos[\chi]\tilde{\rho}_2^2 \\ k_\perp^2\sin[\phi+\chi](\epsilon_{2\parallel}\cos[\phi+\chi]\cos[\chi]+\epsilon_{2\perp}\sin[\phi+\chi]\sin[\chi])-\epsilon_{2\perp}\sin[\chi]\tilde{\rho}_2^2  \\ ik_\perp\epsilon_{2\parallel}\cos[\phi+\chi]\tilde{\rho}_2 \end{array}\right)
\end{eqnarray}
Note that the normalizations on the polarization vectors have been suppressed. We can now use the fact that that the tangential components of the electric and magnetic fields are continuous across the boundary to solve for the reflection coefficients. The matrix elements for $R_1$ are given explicitly in Appendix A. From the reflection coefficients it is obvious that the Casimir force depends on the angle of orientation, $\chi$, which can not be decoupled from the azmuthal angle $\phi$. Thus the integration over $d\mathbf{k}$ in equation \ref{E/A} should be replaced with integrals explicitly over $dk_\perp$ and $d\phi$:
\begin{equation}\label{E/A2}
E/A=\frac{\hbar c}{8\pi^3}\int_0^\infty k_\perp dk_\perp\int_0^{2\pi}d\phi\int_0^\infty\ln(1-\rm{tr}(R_1R_2)e^{-2a\sqrt{\xi_c^2+k_\perp^2}}+\rm{det}(R_1R_2)e^{-4a\sqrt{\xi_c^2+k_\perp^2}})d\xi_c
\end{equation}

Two cases of the above expression are of particular interest, the case where $\epsilon_{i\parallel}=\epsilon_{i\perp}=\epsilon_{i}$ (isotropic case) and the case where $\epsilon_3=\epsilon_{1\parallel}=\epsilon_{2\parallel}=1$ and $\epsilon_{1\perp}=\epsilon_{2\perp}=\infty$ (perfectly anisotropic case). We consider these cases below.
\subsection{Isotropic Case}
It can be shown that in the isotropic limit the reflection matrices given in Appendix A reduce to the isotropic reflection coefficients. In this case the reflection matrices are:
\begin{eqnarray*}
R_1=\left(\begin{array}{cc}
\frac{-\rho_1+\rho_3}{\rho_1+\rho_3}&0\\
0&\frac{\epsilon_3\rho_1-\epsilon_1\rho_3}{\epsilon_3\rho_1+\epsilon_1\rho_3}
\end{array}\right)&,\quad &R_2=\left(\begin{array}{cc}
\frac{-\rho_2+\rho_3}{\rho_2+\rho_3}&0\\
0&\frac{\epsilon_3\rho_2-\epsilon_2\rho_3}{\epsilon_3\rho_2+\epsilon_2\rho_3}
\end{array}\right)
\end{eqnarray*}
the elements of which equal the standard reflection coefficients for the isotropic case. The integrand in equation \ref{E/A} thus reduces to:
\begin{equation}
\ln[1-(r_{1E}r_{2E}+r_{1M}r_{2M})e^{-2\rho_3 a}+(r_{1E}r_{1M}r_{2E}r_{2M})e^{-4\rho_3 a}]
\end{equation}
Here we have dropped the second letter in the subscripts as the matrices are now diagonal. If we let $\epsilon_1=\epsilon_2$,                    ($r_{1E}=r_{2E}$, $r_{1M}=r_{2M}$), Eq. \ref{E/A2} simplifies further to yield the standard Lifshitz expression:
\begin{equation}
E/A=\frac{\hbar c}{4\pi^2}\int_0^{\infty}\!\int_{0}^{\infty}\! k_\perp\ln[(1-r_E^2e^{-2a\rho_3})(1-r_M^2e^{-2a\rho_3})]\  d\xi_c dk_\perp
\end{equation}

\subsection{Perfectly Anisotropic Case}
Another interesting limit of our expression is the totally anisotropic case. In this limit we let $\epsilon_3=\epsilon_{1\parallel}=\epsilon_{2\parallel}=1$ and we will let $\epsilon_{1\perp}$ and $\epsilon_{2\perp}$ go to infinity. The resulting reflection coefficients for Plate 1 are then:
\begin{scriptsize}
\begin{equation}\label{r1ee}
r_{1EE}=\frac{k_\perp^2\sin^2\phi +\xi_c^2(1-2\cos^2\phi)-\sqrt{k_\perp^2+\xi_c^2}\sqrt{k_\perp^2\sin^2\phi +\xi_c^2}}{k_\perp^2\sin^2\phi +\xi_c^2+\sqrt{k_\perp^2+\xi_c^2}\sqrt{k_\perp^2\sin^2\phi +\xi_c^2}}
\end{equation}
\begin{equation}\label{r1em}
r_{1ME}=r_{1EM}=\frac{2 \xi_c \sqrt{k_\perp^2+\xi_c ^2} \cos[\phi ] \sin[\phi ]}{k_\perp^2\sin^2\phi +\xi_c^2+\sqrt{k_\perp^2+\xi_c^2}\sqrt{k_\perp^2\sin^2\phi +\xi_c^2}}
\end{equation}
\begin{equation}\label{r1mm}
r_{1MM}=\frac{-k_\perp^2\sin^2\phi +\xi_c^2(1-2\cos^2\phi)-\sqrt{k_\perp^2+\xi_c^2}\sqrt{k_\perp^2\sin^2\phi +\xi_c^2}}{k_\perp^2\sin^2\phi +\xi_c^2+\sqrt{k_\perp^2+\xi_c^2}\sqrt{k_\perp^2\sin^2\phi +\xi_c^2}}
\end{equation}
\end{scriptsize}
The reflection matrix for Plate 2 is identical to that of Plate 1 with the replacement of $\phi\rightarrow\phi+\chi$. As the interaction energy depends on the product of these matrices, it is demonstrative to see the form of the product under a few simple cases, namely under parallel ($\chi=0$) and perpendicular ($\chi=\pi/2$) alignment.
\begin{eqnarray}
R_1R_2|_{\chi \rightarrow 0,\phi \rightarrow 0}=\left(\begin{array}{cc}
1&0\\
0&\left(\frac{\xi_c-\sqrt{k_\perp^2+\xi_c^2}}{\xi_c+\sqrt{k_\perp^2+\xi_c^2}}\right)^2
\end{array}\right)&,\qquad &R_1R_2|_{\chi \rightarrow 0,\phi \rightarrow \pi /2}=\left(\begin{array}{cc}
0&0\\
0&1
\end{array}\right)
\end{eqnarray}
\begin{eqnarray}
R_1R_2|_{\chi \rightarrow \pi /2,\phi \rightarrow 0}=\left(\begin{array}{cc}
0&0\\
0&\frac{-\xi_c+\sqrt{k_\perp^2+\xi_c^2}}{\xi_c+\sqrt{k_\perp^2+\xi_c^2}}
\end{array}\right)&,\quad &R_1R_2|_{\chi \rightarrow \pi /2,\phi \rightarrow \pi /2}=\left(\begin{array}{cc}
0&0\\
0&\frac{-\xi_c+\sqrt{k_\perp^2+\xi_c^2}}{\xi_c+\sqrt{k_\perp^2+\xi_c^2}}
\end{array}\right)
\end{eqnarray}
From this it is easy to see that for perpendicular alignment the trace of the product of the reflection matrices is reduced with respect to parallel alignment, however it does not go zero for all cases. Thus, one would intuitively predict that the interaction energy will not vanish for perpendicular alignment.

\begin{figure}[h]
\includegraphics[width=4in]{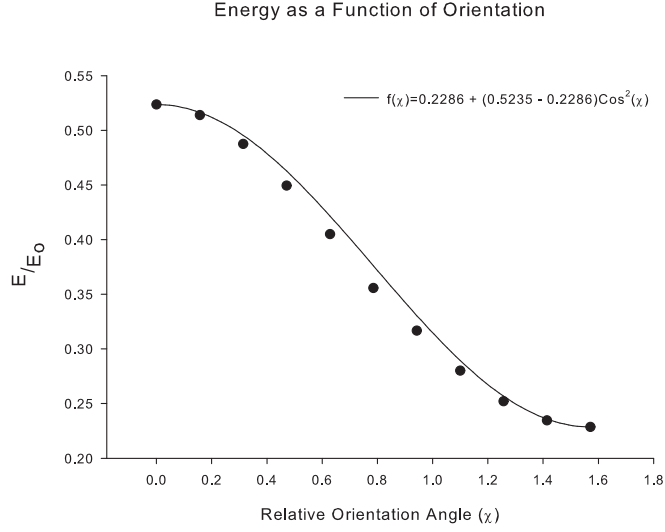}
\caption{\footnotesize{The Normalized Casimir Energy per Unit Area as a function of orientation for a cavity formed with two perfectly anisotropic plates. The values of the energy have been scaled with respect to the (isotropic) perfectly conducting case. The line is an empirical fit: $f(\chi)=0.2286+(0.5235-0.2286)\cos^2[\chi]$ to the points which were calculated using the theory described in this paper. }\label{NewSolution}}
\end{figure}

In order to obtain the energy per unit area as a function of the orientation angle ($\chi$) we have evaluated eq. \ref{E/A2} using the reflection matrices from eqs. [\ref{r1ee}-\ref{r1mm}] and plotted the values in Figure \ref{NewSolution} for a separation of 1 micron. The values have been represented in units of the energy in the Casimir case ($E_0$) and fit to a $\cos^2[\chi]$ function.

The $180^\circ$ symmetry in the orientational dependence could be of particular interest to experimentalists in the field. An experiment could be constructed whereby one of the plates would be rotated relative to the other at a set frequency. The force between the two plates would thus vary at twice the frequency, while all other signals should remain constant or vary at the rotation frequency. Using this method the Casimir force should be easier to isolate from other signals infiltrating such an experiment.The dependence of the energy on the angle of orientation will also produce a torque: $M=\frac{\partial E}{\partial \chi}$. This torque will exhibit the same $2\chi$ dependence as the normal force, however it will be maximum at $\chi=\pi/4$ instead of $\chi=0$. Such a torque would produce a displacement orthogonal to that of the normal force and could in principle be used as another method of measuring the macroscopic effects of the quantum vacuum.

\section{Final Remarks}
In this paper we have developed a general approach to Casimir problems and have used this approach to solve a number of examples. We have shown that this approach yields Casimir's and Lifshitz' expressions in the appropriate limits. Using this approach we have also calculated the Casimir force between anisotropic conductors and shown that this leads to the predicted orientational dependence. It is our hope that we have described this approach in enough detail so that others will find it easy to adopt the formalism in their own calculations.

The authors wish to thank Girish Agarwall for his help and insight. This work has been partially funded by an NSF Grant \#61018 allocated to R.C.

\appendix
\section{Reflection Matrices}
This formalism requires explicitly defined local reflection coefficients, in order to obtain them for boundaries defined by in-plane optical anisotropy we will match the general solutions given in Eqs. \ref{soln3} and \ref{soln12} using Maxwell's equations as the boundary conditions, which state that the tangential components of the electric and magnetic fields must be continuous across the boundary. Explicitly, the boundary conditions for Plate 1 are:
\begin{equation}
\psi_{I,E}\lambda_{Ex}+\psi_{I,-M}\lambda_{-Mx}+\psi_{RE}\lambda_{Ex}+\psi_{R,+M}\lambda_{+Mx}-\psi_{To}\lambda_{ox}-\psi_{Te}\lambda_{ex}=0
\end{equation}
\begin{equation}
\psi_{I,E}\lambda_{Ey}+\psi_{I,-M}\lambda_{-My}+\psi_{RE}\lambda_{Ey}+\psi_{R,+M}\lambda_{+My}-\psi_{To}\lambda_{oy}-\psi_{Te}\lambda_{ey}=0
\end{equation}
\begin{equation}
\psi_{I,E}(\mathbf{k}\times \mathbf{\hat{\lambda}}_{E})_x+\psi_{I,-M}(\mathbf{k}\times \mathbf{\hat{\lambda}}_{-M})_x+ \psi_{RE}(\mathbf{k}'\times \mathbf{\hat{\lambda}}_{E})_x+\psi_{R,+M}(\mathbf{k}'\times \mathbf{\hat{\lambda}}_{+M})_x-\psi_{To}(\mathbf{k}_{1o}\times \mathbf{\hat{\lambda}}_{o})_x-\psi_{Te}(\mathbf{k}_{1e}\times \mathbf{\hat{\lambda}}_{e})_x=0
\end{equation}
\begin{equation}
\psi_{I,E}(\mathbf{k}\times\mathbf{\hat{\lambda}}_{E})_y+\psi_{I,-M}(\mathbf{k}\times\mathbf{\hat{\lambda}}_{-M})_y+\psi_{RE}(\mathbf{k}'\times\mathbf{\hat{\lambda}}_{E})_y+\psi_{R,+M}(\mathbf{k}'\times\mathbf{\hat{\lambda}}_{+M})_y-\psi_{To}(\mathbf{k}_{1o}\times\mathbf{\hat{\lambda}}_{o})_y-\psi_{Te}(\mathbf{k}_{1e}\times\mathbf{\hat{\lambda}}_{e})_y=0
\end{equation}
here $\psi_{i,j}$ is the amplitude of the $i$th wave ($i$ standing for \textbf{I}ncident, \textbf{R}eflected or \textbf{T}ransmitted) with $j$th polarization ($j$ standing for transverse \textbf{E}lectric, transverse \textbf{M}agnetic, \textbf{o}rdinary, or \textbf{e}xtraordinary) and $\lambda_{jk}$ is the $k$th component ($k$ standing for the \textbf{x} or \textbf{y} directions) of the $j$th polarization unit vector. From these four equations, the amplitudes of the reflected waves can be written in terms of the incident waves, convienently arranged in a $2\times 2$ matrix whose elements are given explicitly below.

\begin{equation}
\begin{array}{ll}
r_{1EE}=\left[k_\perp^2 \sin^2[\phi] \rho _3 \left(\rho _3-\tilde{\rho }_1\right) \left\{A+\epsilon_{1\perp} \tilde{\rho }_1^2\right\}+ \rho _1^2 \rho _3 \tilde{\rho }_1 \left\{A+\epsilon_{1\perp} \tilde{\rho }_1^2\right\}\right. +\rho _1 \left\{k_\perp^2 \epsilon_{1\perp} \sin^2[\phi] \rho _3 \left(\rho _3^2-k_\perp^2\right)\right. &\\
\qquad\qquad \left.  -A (k_\perp^2 \sin^2[\phi]+ \cos[2\phi ] \rho _3^2) \tilde{\rho }_1+\epsilon_{1\perp} \rho _3 \left(k_\perp^2-\rho _3^2\right) \tilde{\rho }_1^2-\epsilon_{1\perp} \left(k_\perp^2 \sin^2[\phi]+\cos[2\phi ] \rho _3^2\right) \tilde{\rho }_1^3\right\}&\\
\qquad\qquad\left. -\rho_1^2\left\{k_\perp^2 \cos ^2[\phi] \left(k_\perp^2 (\epsilon_{1\parallel}-\epsilon_{1\perp}) \sin^2[\phi]+\epsilon_{1\perp} \tilde{\rho }_1^2\right)\right.\right. &\\
\qquad\qquad \left.\left. +\rho _3^2 \left(k_\perp^2 \sin^2[\phi] \left(\epsilon_{1\perp} \cos[2\phi]-2\epsilon_{1\parallel}\cos ^2[\phi]\right)+\epsilon_{1\perp} \cos[2\phi ] \tilde{\rho }_1^2\right)\right\}\right]\frac{1}{N_1}&\\
\end{array}
\end{equation}

\begin{equation}
\begin{array}{ll}
r_{1EM}=\frac{2}{N_1} \cos[\phi ] \sin[\phi ] \rho _3 \sqrt{\rho _3^2-k_\perp^2} \left[k_\perp^2 (\epsilon_{1\parallel}-\epsilon_{1\perp}) \sin^2[\phi] \left(k_\perp^2-\rho _1^2\right)+A \rho _1 \tilde{\rho }_1+\epsilon_{1\perp} \left(k_\perp^2-\rho _1^2\right) \tilde{\rho }_1^2+\epsilon_{1\perp} \rho _1 \tilde{\rho }_1^3\right]&\\
\end{array}
\end{equation}

\begin{equation}
\begin{array}{ll}
r_{1ME}=\frac{-2}{N_1} \cos[\phi ] \sin[\phi ] \rho _1 \rho _3 \sqrt{\rho _3^2-k_\perp^2} \left(\rho _1-\tilde{\rho }_1\right) \left[A+\epsilon_{1\perp} \tilde{\rho }_1^2\right] &\\
\end{array}
\end{equation}

\begin{equation}
\begin{array}{ll}
r_{1MM}=\left[k_\perp^2 \sin^2[\phi] \left\{-A \rho _3^2+\epsilon_{1\perp} \rho _1 \rho _3 \left(\rho _3^2-k_\perp^2\right)+\rho _1^2 \left(k_\perp^2 (\epsilon_{1\parallel}-\epsilon_{1\perp}) \cos ^2[\phi]-\left(2\epsilon_{1\parallel} \cos ^2[\phi]-\epsilon_{1\perp} \cos[2\phi]\right) \rho _3^2\right)\right\}\right. &\\
\qquad\qquad +A \left\{k_\perp^2 \sin^2[\phi] \rho _1-\left(k_\perp^2 \sin^2[\phi]-\rho _1^2\right) \rho _3+\cos[2\phi ] \rho _1 \rho _3^2\right\} \tilde{\rho }_1&\\
\qquad\qquad +\epsilon_{1\perp} \left\{-k_\perp^2 \sin^2[\phi] \rho _3^2+\rho _1 \rho _3 \left(k_\perp^2-\rho _3^2\right)+\rho _1^2 \left(k_\perp^2 \cos ^2[\phi]-\cos[2\phi ] \rho _3^2\right)\right\} \tilde{\rho }_1^2&\\
\qquad\qquad\left. +\epsilon_{1\perp} \left\{k_\perp^2 \sin^2[\phi] \rho _1+\left(\rho _1^2-k_\perp^2 \sin^2[\phi]\right) \rho _3+\cos[2\phi ] \rho _1 \rho _3^2\right\} \tilde{\rho }_1^3\right]\frac{1}{N_1}&
\end{array}
\end{equation}

where,
\begin{displaymath}
\begin{array}{ll}
A=-k_\perp^2\left(\epsilon_{1\parallel}-(\epsilon_{1\parallel}-\epsilon_{1\perp}) \sin^2[\phi]\right)&\\
&\\
N_1=\left(\rho _1+\rho _3\right)\left[ k_\perp^2 \sin^2[\phi] \left(\rho _3+\tilde{\rho }_1\right) \left\{A +\epsilon_{1\perp} \tilde{\rho }_1^2\right\}+\rho _1 \left[\epsilon_{1\perp} \rho _3^2 \left\{k_\perp^2 \sin^2[\phi]-\tilde{\rho }_1^2\right\}\right.\right. &\\
\left.\left. \qquad -\rho _3 \tilde{\rho }_1 \left\{A+\epsilon_{1\perp} \tilde{\rho }_1^2\right\}+k_\perp^2 \cos ^2[\phi] \left\{k_\perp^2 (\epsilon_{1\parallel}-\epsilon_{1\perp}) \sin^2[\phi]+\epsilon_{1\perp} \tilde{\rho }_1^2\right\}\right]\right]&\\
\end{array}
\end{displaymath}
Here $r_{1ij}$ represents the amplitude of the reflected wave with $j$ polarization in terms of an incident wave of $i$ polarization at the surface of Plate 1. The reflection coefficients for Plate 2 can obtained by the simple replacement of {$\phi \to \phi+\chi$, $\rho _1\to \rho _2$, $\tilde{\rho }_1\rightarrow\tilde{\rho }_2$, $\epsilon_{1\parallel}\to \epsilon_{2\parallel}$, $\epsilon_{1\perp}\to \epsilon_{2\perp}$}.

\end{document}